\documentclass{iopart}
\usepackage{graphicx}

\usepackage{bm}

\newcommand{\B}{\mbox{\tiny B}}
\newcommand{\nl}{\nonumber \\}

\newcommand{\la}{\langle}
\newcommand{\ra}{\rangle}
\newcommand{\kb}{\rangle\langle}

\newcommand{\Sec}[1]{Sec.\,\ref{#1}}
\newcommand{\be}{\begin{equation}\fl\qquad}
\newcommand{\ee}{\end{equation}}
\newcommand{\bea}{\begin{eqnarray}\fl\qquad}
\newcommand{\eea}{\end{eqnarray}}

\newcommand{\Eq}[1]{Eq.\,(\ref{#1})}
\newcommand{\Eqs}[1]{Eqs.\,(\ref{#1})}
\newcommand{\Fig}[1]{Fig.\,\ref{#1}}
\newcommand{\Figs}[1]{Figs.\,\ref{#1}}

\newcommand{\ind}{{\sf n}}
\newcommand{\bfalp}{\bm\alpha}

\newcommand{\rhonswap}{\rho_{\sf n}^{{ }_{\{\!\rightarrow\!\}}}}
\newcommand{\rhonup}{\rho_{\sf n}^{{ }_{\{\!+\!\}}}}
\newcommand{\rhondown}{\rho_{\sf n}^{{ }_{\{\!-\!\}}}}

\begin{document}

\title[Exact quantum dissipation with driving]
{Exact quantum dissipative dynamics under
external time--dependent fields driving
}

\author{Jian Xu and Rui-Xue Xu\footnote{Author to whom
the correspondence should be addressed.}}
\address{Hefei National Laboratory for Physical Sciences at Microscale,
  University of Science and Technology of China, Hefei, Anhui 230026, China}
\ead{rxxu@ustc.edu.cn}
\author{YiJing Yan}
\address{Department of Chemistry,
  Hong Kong University of Science and Technology, Kowloon,
  Hong Kong SAR, China}
\ead{yyan@ust.hk}
\date{\today}

\begin{abstract}
 Exact and nonperturbative quantum master equation can be constructed
via the calculus on path integral. It results in hierarchical
equations of motion for the reduced density operator.
Involved are also a set of well--defined auxiliary density operators that
resolve not just system--bath coupling strength but also memory.
In this work, we scale these auxiliary  operators individually
to achieve a uniform error tolerance,
as set by the reduced density operator.
An efficient propagator is then proposed
to the hierarchical Liouville--space dynamics
of quantum dissipation.
Numerically exact studies are carried out on the
dephasing effect on population transfer
in the simple
stimulated Raman adiabatic passage
scheme.
We also make assessments on
several perturbative theories
for their applicabilities in the present system of study.

\end{abstract}


\section{Introduction}

 The central problem of quantum dissipation theory is to study
the dynamics of quantum system embedded in quantum thermal bath. The
primary quantity of interest here is the reduced density operator,
$\rho(t)\equiv{\rm tr}_{\B}\rho_{\rm T}(t)$, after the bath degrees
of freedom are all traced out from the total composite
density operator. Due to its fundamental importance, quantum
dissipation theory has remained as an active topic in diversified
fields [1-10]. 
The challenge here, from both formulation and numerical aspects,
is nonperturbative dissipation, with multiple time scales of
memory, under time--dependent external field driving.

 For Gaussian stochastic force,
the influence of bath on system can be characterized by force--force
correlation functions. Exact formalism had then been established via
the Feynman--Vernon influence functional approach [1-5].
Direct numerical integration methods, based on
discretization of the path integral
and summation up of the memory correlated terms,
have been put forward such as
the quasi-adiabatic
propagator method [11-14] or the real-time quantum Monte Carlo scheme [15-19].

The alternative is the differential approach,
especially in a linear form to maximize the numerical advantage.
It has also the advantage in the study
of various dynamics
such as the spectroscopic or control problems \cite{Yan05187}.
The calculus--on--path--integral (COPI) method is hence proposed to construct the
differential counterpart of the
path integral theory,
reported as the hierarchical equations of motion (HEOM)
formalism [21-30].
This formalism can also be
derived via the stochastic description of quantum dissipation [31-35].
The COPI algorithm provides a unified approach to
the influence of quantum environment ensembles,
either canonical or grand canonical,
and either bosonic or fermionic
\cite{Xu07031107,Jin08234703}.
The COPI algorithm also
takes into account the combined effects
of multiple memory time scales, system--bath coupling strengths,
and system anharmonicity.
The resulting  HEOM formalism is therefore
nonperturbative in nature, and always converges in principle.
Moreover, the HEOM formalism is exact, not just for
its propagation equivalent to the path integral theory,
but also for the fact that the initial correlations between
system and bath can now be incorporated by
the steady--state solutions to the HEOM, before
external time--dependent fields taking effect.
Recently, we have further developed a
numerical efficient filtering method for
the propagation of the HEOM \cite{Shi09084105,Shi09164518}.

 In this work, we report a HEOM--based
study on population transfer with dephasing
in the scheme of stimulated Raman
adiabatic passage (STIRAP) \cite{Ber981003}.
The laser control of dissipative systems
has been addressed extensively [39-49],
but mostly on the basis of weak dissipation treatment.
The correlated influence of driving and dissipation
is often important, as demonstrated
previously \cite{Xu046600,Mo05084115}.
With the aid of the numerically exact results, we analyze  the dephasing effects
on transfer dynamics in relation to the STIRAP mechanism
and examine some second--order quantum dissipation theories
for their applicabilities in the systems of study.

The remainder of paper is organized as follows.
We present the HEOM formalism
together with comments on its numerical implementation
in \Sec{theory}, and the derivations in Appendix.
In \Sec{num}, we study the dephasing effect on population
transfer dynamics in the STIRAP scheme.
We report the numerically exact results via the HEOM formalism,
followed by discussions in relation to the STIRAP mechanism.
In \Sec{thcom}, we present the details
of numerical performance of the
HEOM results, and make
concrete assessments on several approximated quantum dissipation theories.
Finally we conclude the paper.

\section{Hierarchical equations of motion formalism for quantum dissipation}
\label{theory}

\subsection{Description of stochastic bath coupling}
\label{thmodel}

  The total system--plus--bath Hamiltonian can be
written  in general as
 \be \label{HT0}
    H_{\rm T} = H(t) + h_{\B}  - \sum_a Q_a \hat F_a .
 \ee
The last term denotes the multi-mode system--bath interactions.
The involving system operators $\{Q_a\}$ are
called the dissipative modes,
through which the generalized Langevin forces
$\{\hat F_a(t) = e^{ih_{\B}t} \hat F_a e^{-ih_{\B}t}\}$
from the bath ($h_{\B}$) act on the system.
For convenience, let the dissipative modes be dimensionless.
The time dependence in the system $H(t)$ arises
from external driving fields.
Throughout this paper, we denote
the inverse
temperature $\beta \equiv 1/(k_{\B}T)$
and set $\hbar \equiv 1$.

 We treat the Langevin forces as Gaussian stochastic processes.
Therefore their effects on the system are completely characterized
by the correlation functions,
 \be \label{FFCorr}
  C_{ab}(t-\tau) = \la \hat F_{a}(t) \hat F_{b}(\tau) \ra_{\B}.
 \ee
Here,
$\la\hat O\,\ra_{\B}\equiv{\rm tr}_{\B}(\hat O \rho^{\rm eq}_{\B})$
denotes the thermodynamics average over the canonical
ensembles of the bosonic bath.
 The correlation functions satisfy the symmetry and
 detailed--balance relations, or equivalently
the fluctuation--dissipation theorem \cite{Wei08,Yan05187}:
 \be \label{FDT}
  C_{ab}(t) = \frac{1}{\pi} \int_{-\infty}^{\infty}\!d\omega\,
     \frac{e^{-i \omega t}J_{ab}(\omega)}
     {1-e^{-\beta\omega}}\, ,
 \ee
with $J_{ab}(\omega)= - J_{ba}(-\omega) = J_{ba}^{\ast}(\omega)$
being the bath spectral density functions.
 The HEOM formalism requires $C_{ab}(t)$
be expanded in certain series form, so that the hierarchy can be constructed via
consecutive time derivatives on path integral.
Various schemes \cite{Tan906676,Xu07031107,Mei993365,Xu037}
have been proposed to expand $C_{ab}(t)$
in exponential series,
on the basis of
analytical continuation evaluation of \Eq{FDT}.
In particular, the hybrid scheme
that also exploits quadrature integration method
is applicable for arbitrary spectral density functions  \cite{Zhe09164708}.

  For simplicity we set $C_{ab}(t)=C_{aa}(t)\delta_{ab}$.
In this case the contributions from different dissipative modes $\{Q_a\}$
are additive. Without loss of generality,
we present the formalism explicitly
only for the single--dissipative--mode case, $Q_a=Q$.
We thus omit the index $a$ for clarity of formulation.
 We also adopt the super-Drude model,
\be\label{Jmodel}
  J(\omega) = \frac{\eta\omega}{[(\omega/\gamma)^2+1]^2} \,.
\ee
The corresponding correlation function can be analytically
evaluated as \cite{Yan05187,Xu07031107,Xu037}
\be\label{Catexpan}
  C(t\geq 0)= [\nu  + (\bar\nu_r+i\bar\nu_i)\,\gamma t] e^{-\gamma t}
  + \sum_{m=1}^{M} \check\nu_m e^{-\check\gamma_mt}+\delta C(t).
\ee
All coefficients here are real and given in Appendix
[cf.\ \Eqs{nucoefapp} and (\ref{checknu})].
The first term arises from pole of
the spectral density function, which is of rank two.
The second term is from the Matsubara poles, with $\check\gamma_m \equiv 2\pi m/\beta$
being the Matsubara frequency.
The last term is the Matsubara residue, which
would approach to zero if $M\rightarrow\infty$.
In this work, we adopt the Markovian residue ansatz \cite{Ish053131,Tan06082001}, i.e.,
$\check\gamma_m e^{-\check\gamma_m t}\big|_{m>M} \approx \delta(t)$;
thus,
\be\label{detC}
 \delta C(t) \simeq \Delta\, \delta(t); \quad \ \
 \Delta =\!\!\sum_{m=M+1}^\infty  \frac{\check\nu_m}{\check\gamma_m}
 = \frac{\eta}{\beta}- \frac{\nu+\bar\nu_r}{\gamma}
  - \sum_{m=1}^{M}\frac{\check\nu_m}{\check\gamma_m}\,.
\ee

\subsection{The HEOM formalism}
\label{theom}

 The dynamics quantities in the HEOM formalism
are the reduced density
operator $\rho(t)$ and
a set of  auxiliary density operators (ADOs),
$\{\rho_{\ind}(t)\}$, that hierarchically resolve
the memory contents of the bath correlation functions
in the exponential series expansion of \Eq{Catexpan}.
The index ${\ind}$
that specifies an $N^{\rm th}$--tier ADO $\rho_{\ind}$
consists of a series of nonnegative integers,
\be \label{indn}
 {\ind}\equiv
   \left\{n,n',\bar n,\bar n',
   \check n_1,\cdots,\check n_M \right\},
\ \  {\rm with} \ \
 n+n'+\bar n+\bar n'+\check n_1+\cdots+\check n_M=N .
\ee
Comparing to the reduced density
operator $\rho(t) \equiv \rho_{\sf 0}(t)$ of primary interest,
the specified $\rho_{\ind}$ would have
the order of
$|\nu|^{n+n'}\!\cdot\!|\bar\nu_r+i\bar\nu_i|^{\bar n+\bar n'}
\!\cdot\!\prod_{m=1}^M |\check\nu_m|^{\check n_m}$,
for its dependence on the individual components of interaction bath correlation functions
in the series expansion of \Eq{Catexpan}.
These scaling factors will be
incorporated properly in the final dimensionless $\rho_{\ind}$,
in order to validate a filtering algorithm for the numerical efficiency
of the HEOM formalism.
On the other hand,
the indices in the set $\ind$ of \Eq{indn}
cover all accessible derivatives of the Feynman--Vernon influence functional;
see Appendix for the details.

 The final HEOM formalism is summarized as follows. It has the generic form of
\be \label{dotrhon}
 \dot\rho_{\ind} =-[i{\cal L}(t) +{\Gamma_{\ind}} + \delta\mathcal{R}]\rho_{\ind}
  +\rhonswap +\rhondown + \rhonup .
\ee
Here, ${\cal L}(t)\rho_{\ind}\equiv[H(t),\rho_{\ind}]$, which depends in general
on the external driving fields;
$\Gamma_{\ind}$ is the damping parameter that collects all related exponents,
and
$\delta\mathcal{R}$ is the residue dissipation superoperator due to
$\delta C(t)$. For the bath correlation
function in the series expansion, \Eq{Catexpan} with \Eq{detC},
they are given respectively by
\be\label{Gam_dR}
   {\Gamma}_{\ind}\equiv(n+n'+{\bar n}+{\bar n}')\gamma
     +  \sum_{m=1}^M \check n_m \check \gamma_m \,  ,
\quad\  \delta\mathcal{R}\rho_{\ind} =  \Delta [Q,[Q,\rho_{\ind}]]\,.
\ee
Apparently, $\Gamma_{\sf 0} \equiv {\Gamma}_{\ind}|_{N=0} = 0$.

 The last three terms in \Eq{dotrhon} denote how
the specified $N^{\rm th}$--tier ADO $\rho_{\ind}$ depends on other ADOs of the
same tier, the $(N-1)^{\rm th}$--tier,
and the $(N+1)^{\rm th}$--tier, respectively.
For the bath correlation function in \Eq{Catexpan},
they are given explicitly by
\bea
 \rhonswap = \vec{\lambda}^{\,r}_n {\rho}_{\vec{\ind}}
             + \vec{\lambda}^{\,i}_{n'} {\rho}_{\vec{\ind}'}   \, ,
\nonumber \\ 
\fl\qquad
  \rhondown =-i [Q,\lambda_n   {\rho}_{{\ind}^{-}}]
             + \{Q,\lambda_{n'}{\rho}_{{\ind}^{\prime-}}\}
             -i \Big[Q,\sum_{m=1}^M \check\lambda_{\check n_m}
                        {\rho}_{{\check{\ind}}_{m}^{-}}
                \Big]  \, ,
\label{rhonasso} \\ 
\fl\qquad
  \rhonup = -i
  \Big[Q,
        \lambda_{n+1}{\rho}_{{\ind}^{\!+}}
      + {\rm sign}(\bar\nu_r)\!\cdot\!\bar\lambda^{\,r}_{\bar n+1}
           {\rho}_{\bar{\ind}^{\!+}}
      - \bar\lambda^{\,i}_{\bar n'+1} {\rho}_{\bar{\ind}^{\prime\!+}}
      - \sum_{m=1}^{M} \check\lambda_{\check n_m+1}
           {\rho}_{{\check{\ind}}^{\!+}_{m}}
   \Big] \, .
\nonumber
\eea
Here, $\lambda_n  = \sqrt{n|\nu|}$\,,    
$\check\lambda_{\check n_m} = \sqrt{\check{n}_m |\check\nu_m|}$\,,
  $\bar\lambda^{\,r/i}_n = \sqrt{n|\bar\nu_{r/i}|}$\,,
and $\vec{\lambda}^{\,r/i}_n  = \gamma\bar\lambda^{\,r/i}_{n+1}  \sqrt{\bar n /|\nu|}$\,,
with the italic--font {\it indices} being from those in $\ind$ of \Eq{indn}.
 The indexes variations in \Eq{rhonasso} that specify
those ADOs participating in the equation of $\dot\rho_{\ind}$ are exemplified as follows:
\be\label{sfnvar}
\vec{\ind} \equiv
   \left\{n+1,n',\bar n-1,\bar n',
   \check n_1,\cdots,\check n_M \right\},
\quad
 \ind^{\pm} \equiv \left\{n\pm 1,n',\bar n,\bar n',
   \check n_1,\cdots,\check n_M \right\}.
\ee
Similarly,
$\vec{\ind}'$ differs from $\ind$ of \Eq{indn} only by
changing $(n', \bar n')$ to $(n'+1, \bar n'-1)$,
while
$\check{\ind}_{m}^\pm$ by changing $\check n_m$ to $\check n_m\pm 1$,
and so on.
Also note that ${\rho}_{\vec{\ind}}$ is an $N^{\rm th}$--tier ADO,
while ${\rho}_{{\ind}^{\!\pm}}$ is of an $(N\pm 1)^{\rm th}$ tier,
as inferred from the second identity of \Eq{indn}.

 The initial conditions to the HEOM
in the study of driven dissipative dynamics
are obtained via the steady--state solutions to \Eq{dotrhon},
before the time--dependent external fields interactions.
For the steady--state solutions satisfying $\dot\rho^{\rm st}_{\ind}=0$,
\Eq{dotrhon} reduces to a set of linear equations,
under the constraint of ${\rm Tr}\rho_{\sf 0} = 1$.
The resulting $\rho^{\rm st}_{\ind}$ is used as
the initial $\rho_{\ind}(t_0)$ to the HEOM.
The initial system--bath correlations
are accounted for by
those nonzero initial ADOs.


\subsection{Comments on numerical implementation}
\label{theoryC}

 For the numerical HEOM propagation, we would like to have certain convenient working
index scheme to track the multiple indices, denoted
now as an {\it ordered set} of $\ind = \{n_1,\cdots,n_K\}$, that specifies $\rho_{\ind}$.
Here we will provide two such schemes.
The number of the $N^{\rm th}$--tier ADOs, with $n_1+\cdots + n_K = N$,
is $\frac{(N+K-1)!}{N!\,(K-1)!}\equiv\left[\,^N_K\,\right]$.
In one scheme, the ADOs are arranged as
$\rho_{\ind}\equiv\rho_{j_{\ind}}$ with $j_{\ind}$ initialized by
$j_{{\ind}={\sf 0}}\equiv 0$ and then
\be\label{appindex}
  j_{{\ind}\neq{\sf 0}} = j_{n_1\cdots n_K}
=   \sum_{N'=0}^{N-1}   \left[\,^{N'}_K\,\right] +
    \sum_{k=1}^K \sum_{q=0}^{N-s_k}
    \left[\,^{\ \  q}_{K-k}\,\right];
    \    \     s_k = 1 + n_1 + \cdots + n_k .
\ee
Let $L$ be the maximum level of the hierarchical tier.
The total number of the ADOs $\{\rho_{\ind}; 0\leq N \leq L\}$ is
\be \label{calN}
 {\cal N} = \sum_{N=0}^{L} \left[\,^N_K\,\right]
    \equiv \left\{\,^L_K\,\right\} .
\ee
In another scheme, ADOs
can also be arranged as $\rho_{\ind}\equiv\rho_{l_{\ind}}$;
$l_{\ind}=0,\cdots, {\cal N}-1$,
with
\be\label{totindex}
  l_{\ind} = l_{n_1\cdots n_K}
=   n_1 + \sum_{k=2}^K \sum_{q=1}^{n_k}
  \left\{\,^{L+q-(n_{k}+\cdots + n_K)}_{\qquad k-1}\,\right\}.
\ee
Both schemes, \Eq{appindex} and \Eq{totindex}, allow easy tracking of the
coupled ADOs in the HEOM.
The former [\Eq{appindex}] is somewhat more convenient in the filtering propagator
described soon since it does not depend on $L$.

 The major difficulty in implementing the HEOM formalism
is its numerical tractability.
The number of ADOs, ${\cal N}$ of \Eq{calN}, itself alone
can be huge in the case of strong non-Markovian
system--bath coupling and/or low temperature,
as large $L$ and/or large $K$ implied.
Thus, a brute--force implementation
is greatly limited by the memory and central processing
unit (CPU) capability of computer facility, even for a two--level system
where each ADO is a $2\times 2$ matrix.

  To facilitate this problem, Shi, Xu, Yan and coworkers
have recently proposed an efficient numerical
filtering algorithm that often reduces
the effective number of ADOs by order of magnitude \cite{Shi09084105,Shi09164518}.
In reality, there is usually only a very small fraction of
total ADOs significant to the reduced system dynamics.
To validate the accuracy--controlled numerical filtering algorithm,
the present HEOM formalism has been scaled properly so that
all ADOs $\{\rho_{\sf n}(t)\}$ are of a uniform error tolerance.
This remarkable feature
is suggested by comparing the HEOM theory
with the stochastic bath interaction field approach
in the case of Gaussian--Markovian dissipation \cite{Shi09164518}.
The involving ADOs are just the expansion coefficients,
over the normalized harmonic wave functions that
are used as the basis set
for resolving the diffusive bath field \cite{Shi09164518}.
Our numerical HEOM propagator
exploits the filtering algorithm \cite{Shi09084105}.
It goes simply as follows. If a $\rho_{\ind}(t)$
whose matrix elements amplitudes
become all
smaller than the pre--chosen error tolerance,
it is set to be zero.
Apparently, the filtering algorithm also automatically
truncate the required hierarchy
level {\it on--the--fly} during numerical
propagation.
By far the truncation for the Matsubara expansion
still goes by checking convergency.

\section{Effect of dephasing on population transfer via STIRAP}
\label{num}
\subsection{Numerical results}

\begin{figure}
\begin{center}
\includegraphics[width=1.0\columnwidth]{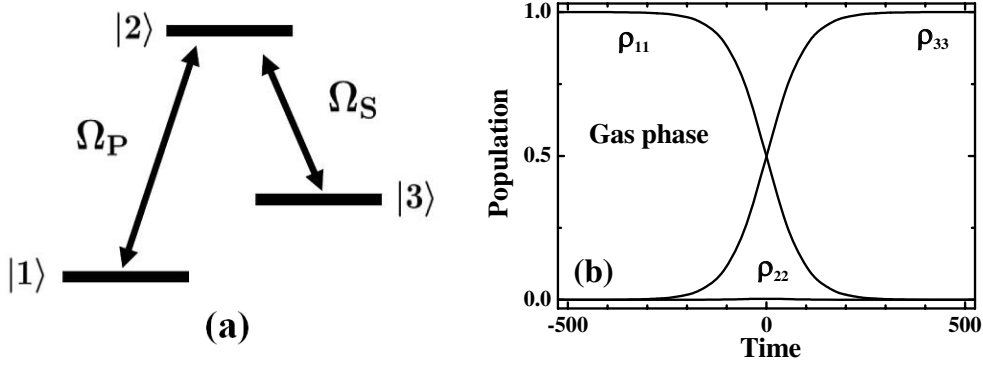}
\caption{(a) A schematic view of the STIRAP of a three-level $\Lambda$  system.
(b) Population transfer under the STIRAP
scheme for the dissipation-free gas phase.
The time is in the unit of $\beta$.
See parameters in the text.}
\label{fig1}
\end{center}
\end{figure}

   The STIRAP is celebrated as an efficient and robust method
for population transfer \cite{Ber981003}.
It is characterized by its counterintuitive field configuration.
For a three-level ${\Lambda}$--system as \Fig{fig1},
the Stokes pulse proceeds the pump pulse, and
the intermediate state remains effective in dark.
The STIRAP mechanism  \cite{Ber981003} is rooted at the existence
of the coherent population trapping state under the two--photon
resonance condition, $\omega_S-\omega_P = \epsilon_1 - \epsilon_3$,
in the ${\Lambda}$--system.
Dephasing destroys this condition in terms of
resonance and/or the existence of coherent population trapping state.
The effect of dephasing on the simple STIRAP scheme
has been studied extensively, but with approximations.
These include phenomenological/perturbative methods
[44-46],
or classical/stochastical bath treatments
[47-49].

 We revisit the dephasing effect on the
simple STIRAP--based population transfer,
as the exact dissipative dynamics are now established
with the present HEOM formalism.
We also examine
three schemes of second--order approximation [20, 52-55]:
(i) The Redfield theory, which neglects the correlated driving--and--dissipation effect;
(ii) CS--COP, which is the conventional time--nonlocal quantum master equation,
 including the field--dressed dissipation contribution,
  and equivalent to the present HEOM truncated at the first tier;
(iii) CODDE,
  in which the driving field--free part of dissipation superoperator is time--local,
  while the field--dressed part is time--nonlocal.
  Neglecting the latter leads it to the Redfield theory.

 The total Hamiltonian
under the rotating wave approximation
assumes
\be\label{HM_DBOa}
  H_{\rm T}(t) = \Omega_P(t)\hat D_P+\Omega_S(t)\hat D_S
  +\sum_j\Bigl[\frac{p_j^2}{2m_j}
      +\frac{1}{2} m_j\omega_j^2
         \Bigl(x_j-\sum_{a}
         \frac{c_{aj}Q_a}{m_j\omega_j^2}\Bigr)^2\Bigr].
\ee
Here, $\hat D_P\equiv|1\kb 2|+|2\kb 1|$
and $\hat D_S\equiv|2\kb 3|+|3\kb 2|$,
while $\Omega_P(t)$ and $\Omega_S(t)$
denote the Rabi frequencies of the resonant pump and Stokes fields,
respectively. The dissipative mode
$Q_a = |a\ra\la a|$ is responsible for dephasing.
The interaction spectral density function $J_a(\omega)\equiv (\pi/2) \sum_j
[c^2_{aj}/(m_j\omega_j)]\delta(\omega-\omega_j)$ assumes
super--Drude as \Eq{Jmodel}.
The system Hamiltonian is then
\be \label{Htnum}
   H(t) = \Omega_P(t)\hat D_P+\Omega_S(t)\hat D_S
   + \sum_{a=1}^3\delta \epsilon_a|a\kb a|\, .
\ee
The Caldeira--Leggett renormalization energy \cite{Cal83374,Cal83587}
is $\delta\epsilon_a=\frac{1}{\pi}\int_{0}^{\infty}\!\!d\omega\,J_{a}(\omega)/\omega
= \frac{1}{4}\eta_a\gamma_a$,
for  the super--Drude model [\Eq{Jmodel}].
In  the STIRAP configuration,
it would relate to the effective detuning at short--time
of the pump or Stokes field, as inferred
from the analytical result of driven Brownian oscillator \cite{Yan05187,Xu09074107}.
%

  We set the pump and Stokes fields be of same Gaussian shape,
$\Omega_P(t+t_P) = \Omega_S(t+t_S) = A \exp[-\frac{1}{2}(wt)^2]$,
but center them at $t_P=200\,\beta$ and $t_S=-200\,\beta$,
respectively and counter-intuitively.
The driving strength and inverse duration parameters
are set to be $\beta A=0.1$ and $\beta w=0.005$.
The corresponding dissipation--free transfer dynamics
is shown in \Fig{fig1}(b).
As here the bath influence is considered to be pure-dephasing
in the absence of fields, the initial system
is just chosen to be completely on the $|1\ra$ state
and all the ADOs are zero.
For the effect of bath, we set
the coupling strength $\eta = 0.64$ [cf.\ \Eq{Jmodel}],
and consider both the Markovian and non-Markovian cases
as follows.

\begin{figure}
\begin{center}
\includegraphics[width=1.0\columnwidth]{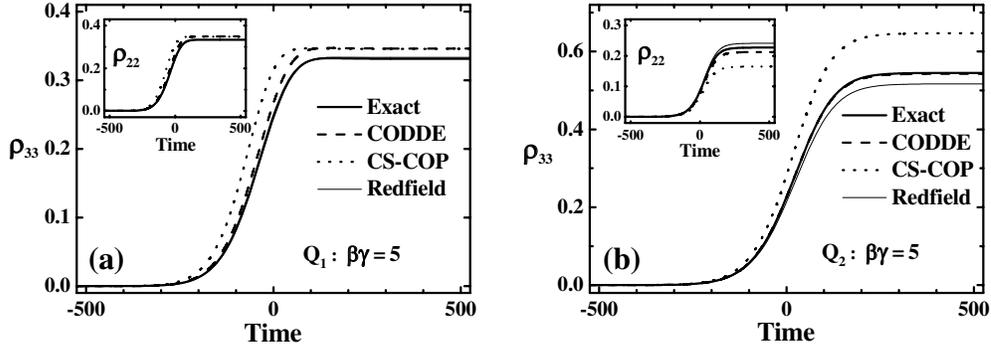}
\caption{Evolutions of $\rho_{33}$ and
$\rho_{22}$ (in insets) via the exact HEOM (solid),
CODDE (dash), CS-COP (dot) and the Redfield equation (thin-solid)
for single--dissipative--mode case: (a)
$Q_1 = |1\ra\la1|$ and (b) $Q_2 = |2\ra\la2|$.
The system--bath coupling strength $\eta=0.64$.
The parameter $\beta\gamma = 5$ exemplifies the Markovian condition.
The time is in the unit of $\beta$.}
\label{fig2}
\end{center}
\end{figure}

\begin{figure}
\begin{center}
\includegraphics[width=1.0\columnwidth]{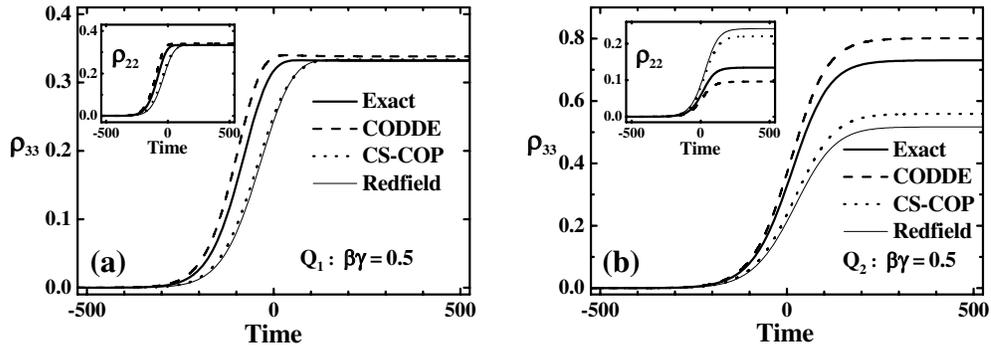}
\caption{Same as \Fig{fig2}, except the parameter
$\beta\gamma = 0.5$ exemplifies  the non-Markovian condition.
The time is in the unit of $\beta$.}
\label{fig3}
\end{center}
\end{figure}

\begin{figure}
\begin{center}
\includegraphics[width=1.0\columnwidth]{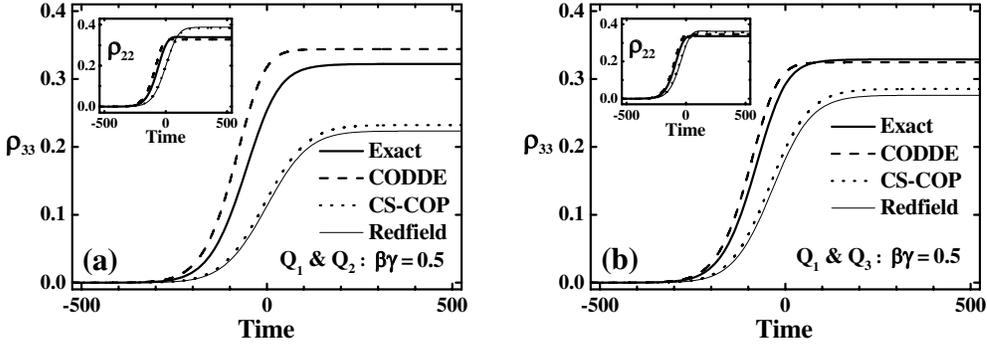}
\caption{Two-mode dissipation cases: (a) $Q_1 = |1\ra\la1|$ and $Q_2 =
|2\ra\la2|$ together, and (b) $Q_1 = |1\ra\la1|$ and  $Q_3 = |3\ra\la3|$ together,
evaluated with $\beta\gamma = 0.5$.
The system--bath coupling strength $\eta = 0.64$.
The time is in the unit of $\beta$.}
\label{fig4}
\end{center}
\end{figure}

 The Markovian transfer dynamics, under the influence of single dephasing mode
of either
$Q_1 = |1\ra\la1|$ or $Q_2 = |2\ra\la2|$,
is exemplified in \Fig{fig2},  with
$\beta\gamma = 5$.
We observe:
({\bf i}) The $Q_1$-mode effect shown in \Fig{fig2}(a) leads to
all three populations about $1/3$ after the driving;
({\bf ii}) The $Q_2$-mode effect shown in \Fig{fig2}(b)
is less sensitive than its $Q_1$ counterpart, achieving
a higher transfer efficiency, despite it is only about 0.55.

The non--Markovian transfer dynamics is exemplified in \Fig{fig3}, with
$\beta\gamma = 0.5$.
In comparison with the Markovian counterparts, we observe:
(${\bf iii}$) The $Q_1$-mode case behaves about the same;
(${\bf iv}$) But the $Q_2$-mode results in a higher transfer yield,
increasing to about 0.73 via the exact calculation.
 We have also calculated the influences of $Q_3=|3\ra\la3|$
for both Markovian and non-Markovian cases.
The results (not shown here) are similar to those of $Q_1$,
except for some small oscillations.

  Two double--modes ($Q_1+Q_2$ and $Q_1+Q_3$ uncorrelated) non--Markovian dephasing
dynamics are shown in \Fig{fig4}(a) and (b), respectively.
They are insensitive to the non--Markovian parameter,
and both reach at the final equal--populations,
based on the numerically exact results.
Comments on the approximated schemes,
the CODDE, CS--COP, and Redfield theory,
presented in \Figs{fig2}--\ref{fig4}
will be given later; see \Sec{thcom2}.

\subsection{Discussions}

 The above observations can be understood by the
well--established STIRAP mechanism \cite{Ber981003}.
The $Q_1$--mode, which associates with the fluctuation of
level $|1\ra$, easily destroy the two--photon
resonance (TPR) condition, as described at the beginning of \Sec{num}.
Thus, it ends up with the observed equal populations in all accessible levels
by the strong fields, as consistent with the
analysis in \cite{Iva04063409}.
The similarity between the $Q_3$ and $Q_1$ influences is also
explained.
The same reason accounts further for the case of
uncorrelated
two modes ($Q_1+Q_2$ or $Q_1+Q_3$) dephasing, as depicted  in \Fig{fig4}.
It is anticipated that when $\gamma \ll w$
(termed as the {\it linear adiabatic limit} below),
the equal--population will be broken to be
in favor of $|3\ra$,
due to the marginally partial fulfilment of the TPR condition.

 On the other hand,
the $Q_2$-mode is associated with
the fluctuation of the intermediate level
$|2\ra$. It alone does not affect the TPR condition.
However, this condition, based on the numerically exact results shown in this work,
is not sufficient to retain the coherent
population trapping  state,
chosen {\it ad hoc} earlier for the
dephasing--free STIRAP scenario in \Fig{fig1}(b).
It is anticipated that
the coherent population trapping state may be recovered
in the aforementioned linear adiabatic limit.
This is in line with the observation-(${\bf iv}$),
where the non--Markovian population transfer with
single $Q_2$-mode dephasing [\Fig{fig3}(b)]
is of higher efficiency than its
Markovian counterpart [\Fig{fig2}(b)].
The previous study based on perturbative dephasing dynamics \cite{Iva04063409}
has also shown that the single $Q_2$-mode does not
affect the transfer efficiency in
the linear adiabatic limit.
Nevertheless, STIRAP in the presence of complex dephasing,
if 100\% transfer ever achievable, would require dynamics feedback control
of pump or Stokes laser frequency \cite{Che06035601}.
This would involve chirp and realize STIRAP
in a nonlinear adiabatic condition,
rather than the linear simplification considered here.

\section{Assessments on theoretical methods and concluding remarks}
\label{thcom}

\subsection{Numerical performance of the HEOM formalism}
\label{thcom1}

\begin{table}
\caption{\label{Tab1}  Performance of HEOM formalism with filtering.}
\begin{indented}
\lineup
\item[]
\begin{tabular}{cccccc}
\br
$10^{-6}$--filter
           &CPU\,(min)
                &${\cal N}_{\rm max}$ {\tiny(${\cal N}$)}
                                                &$L_{\rm max}$
                                                    &$\kappa$
                                                          &$\check\kappa_{1}$
\cr\mr
Fig.\,2(b) &932 &$19765$ {\tiny($8.44\times10^6$)} &17 &0.95 &1.28
 \cr\mr
Fig.\,3(b) &15  &$400$ {\tiny($9.24\times10^4$)}   &9 &1.24 &348
 \cr\mr
Fig.\,4(b) &266 &$3664$ {\tiny($1.00\times10^7$)}  &9 &1.24 &348
 \cr
\br

\end{tabular}
\end{indented}
\end{table}

 The numerical performance of the HEOM formalism with filtering
is summarized in Table \ref{Tab1}, for the systems
reported in the three figures' (b)-panels.
The CPU time is for a single Intel(R) Xeon(R) processor@3.00GHz
to calculate the exact result in each (b)-panel for the time period
$-1200\,\beta < t < 2000\,\beta$ with the time step $dt=0.01\beta$
using the fourth-Order Runge-Kutta propagator;
${\cal N}_{\rm max}$ denotes the largest number
of active ADOs and
$L_{\rm max}$ the highest tier level, ever survived
in the entire time span of the numerical propagation.
The filtering error tolerance is chosen to be
10$^{-6}$, following our previous work \cite{Shi09084105}.
We input $M=6$ for the number of
Matsubara terms being explicitly included,
which has been tested to give converged results of $\rho(t)=\rho_{\sf 0}(t)$  in all calculations.
The total number ${\cal N}$ of mathematical ADOs
follows \Eq{calN} and is given inside the parentheses.
The effect of filtering is clearly seen.
The number of active ADOs with filtering
is insensitive to the  input $M$, as long as
it is large enough.
In the present study, the number of active ADOs reaches ${\cal N}_{\rm max}$
only during the period about $-250\,\beta < t < 500\,\beta$
and grows up or drops down dramatically outside that period
with the fields turning on or getting over.
Apparently, ${\cal N}_{\rm max}$
increases with the number of dissipative modes.

At least one $(L_{\rm max})^{\rm th}$--tier ADO
actively participates during the HEOM propagation.
Its leading contribution to the reduced density operator
is of $(2L_{\rm max})^{\rm th}$ order in the
system--bath interaction.
Physically, ${L}_{\rm max}$ is closely related
to the modulation $\kappa$-parameter \cite{Xu05041103},
introduced originally by Kubo for
motional narrowing problem \cite{Kub85}.
This dimensionless parameter is
determined via $\kappa \equiv \gamma/\sqrt{\nu}$, or similar,
for each individual
exponential component in \Eq{Catexpan}.
The last two columns of Table \ref{Tab1} are
the modulation parameters $\kappa$ and
$\check\kappa_{m=1}$
of the leading Matsubara term.
The modulation $\kappa$-parameter relation to
the value of $L_{\rm max}$ \cite{Xu05041103} can be clearly seen.
In both the Markovian and non-Markovian cases of the present study,
$\check\kappa_m=\check\gamma_m/\sqrt{|\check\nu_m|}$
monotonically increases with $m$, cf.\ \Eq{checknu}.
Actually, the Matsubara series truncation $M$ in \Eq{Catexpan}
can be estimated via its
reaching the fast modulation condition, $\check\kappa_{M}\gg 1$.
As the temperature decreases, $\check\kappa_m$
getting smaller, and eventually cause the
value of $L_{\rm max}$  be pretty large.
The present HEOM construction is
based on the Matsubara series expansion,
which may no longer be numerically implementable in the
extremely low temperature regime.
Alternative expansion method such as the hybrid scheme \cite{Zhe09164708}
is needed to the required HEOM construction.

\subsection{Assessments on three second--order approximated theories}
\label{thcom2}

  With the exact results, we can now make concrete assessments
on the three second--order
approximated schemes, the Redfield theory, CS--COP, and CODDE,
exploited in the numerical demonstrations.
The dissipative modes
$\{Q_a=|a\kb a|\}$  considered in \Sec{num} are all of pure dephasing,
in the absence of external fields.
The Redfield theory would be exact if there were
no correlated driving--and--dissipation
effect \cite{Yan05187,Yan002068}.
Therefore, the non--Markovian dynamics
manifest here as the correlated driving and dissipation.
Apparently, the Redfield theory, by its Markovian nature,
is independent of the
width $\gamma$--parameter of bath spectral density.
Observed is also the fact that the schemes of approximation are sensitive to
the $Q_2$-mode rather than the $Q_1$- or $Q_3$-mode dephasing.
This fact is also easily understood,
by considering their relations to the STIRAP mechanism
as discussed earlier.
Further remarks on the approximated theories for their applicabilities
in the systems of study are as follows.

 The CS--COP theory \cite{Yan05187,Xu037,Xu029196} is overall most unsatisfactory,
despite it contains formally a description of
memory and driving--and--dissipation correlation.
Even in the non--Markovian $Q_2$-mode case of
\Fig{fig3}(b) where it appears to be superior than the Redfield theory,
the CS--COP results in a  {\it decreased} transfer efficiency
from its Markovian counterpart shown in \Fig{fig2}(b).
This is qualitatively
contradictory to the physical anticipation, as discussed earlier.

 The CODDE \cite{Yan05187,Xu037,Xu029196} appears to be the most
favorable perturbation theory.
It gives the best approximated transfer dynamics in all cases presented in \Sec{num},
except the one to be discussed soon.
Its overall superiority is also true in the driven Brownian oscillator systems
\cite{Yan05187,Mo05084115}.
The CODDE is actually a modified Redfield theory, with inclusion
of correlated driving--and--dissipation effects.
The involving field--dressed dissipation kernel is time--nonlocal
but constructed with a partial ordering resummation, rather than
the chronological ordering prescription
that characterizes the CS--COP \cite{Yan05187,Xu037,Xu029196}.

The only exception is the Markovian
$Q_1$--mode case shown in \Fig{fig2}(a),
where the Redfield dynamics is almost exact.
The reason for this exception is also accountable.
As we mentioned earlier, the non--Markovian dynamics
manifest as the correlated driving and dissipation.
This correlated effect
diminishes in both
the fast-- and slow--modulation regimes,
as inferred from   the exact and analytical results of
driven Brownian oscillator systems \cite{Xu09074107}.
This conclusion can be carried over to the present system of study, as suggested here.
Apparently, the identical  value of $\beta\gamma=5$,
adopted in the two cases of \Fig{fig2},
acquires the fast--modulation limit
for the  $Q_1$--mode, but not yet
for the $Q_2$--mode. In the latter
case, the CODDE resumes its superiority.

\subsection{Closing remarks}
\label{sum}

  In summary, we have presented a hierarchical Liouville--space
approach, which is exact and also quite tractable numerically,
to general quantum
dissipation systems under external driving fields.
The auxiliary
density operators are all
of a uniform error tolerance,
as that of the reduced density operator.
We also make comments on the numerical
facilitation about the multiple--index assignment and the filtering
algorithm.
We numerically study the dephasing effects on the
population transfer, with a fixed simple STIRAP configuration, and
present a concrete assessment on various approximation schemes.

\ack
 Support from the National Natural Science Foundation of China
 (20533060 and 20773114),
 National Basic Research Program of China (2006CB922004),
 and RGC Hong Kong (604007 and 604508) is acknowledged.

\appendix

\section*{Appendix. Construction of HEOM via the COPI approach}
\setcounter{section}{1}
\label{thappA}

This appendix gives the details of the COPI approach to the HEOM
formalism. It starts with the influence functional in
 path integral.
 Let $\{|\alpha\ra\}$ be a basis set in the system subspace
 and set ${\bm \alpha}\equiv(\alpha,\alpha')$ for abbreviation.
Denote the evolution of reduced
 system density operator in the $\alpha$-representation by
\be \label{rhotPI_def}
 \rho({\bm \alpha},t) \equiv \rho(\alpha,\alpha'\!,t)
 \equiv \!\int\!d{\bm \alpha}_0\, {\cal U}({\bm \alpha},t;{\bm \alpha}_0,t_0)
 \rho({\bm \alpha}_0,t_0).
\ee
Here, the reduced Liouville-space propagator is
 \be \label{calGPI}
   {\cal U}(\bfalp,t;\bfalp_0,t_0)
 = \int_{\bfalp_0[t_0]}^{\bfalp[t]}   \!\!  {\cal D}{\bfalp} \,
     e^{iS[\alpha]} {\cal F}[\bfalp] e^{-iS[\alpha']}.
 \ee
The effects of bath on the reduced system are contained completely in
the influence functional ${\cal F}$.
For Gaussian stochastic forces $\{\hat F_{a}(t)\}$
from fluctuating bath,
it assumes the Feynman--Vernon form \cite{Fey63118},
which can be recast as [27-29]:
\be \label{FV_FPhiA}
   {\cal F}[\bfalp] =
      \exp\Bigl\{-\int_{t_0}^t\!\!d\tau\,\sum_{a} {\cal A}_a[\bfalp(\tau)]\,
   {\cal B}_{a}(\tau;\{\bfalp\})\Bigr\},
 \ee
with
\be \label{calQAB}
  {\cal A}_a[\bfalp(t)] =
    Q_a[\alpha(t)]-Q_a[\alpha'(t)],
\ee
\be \label{ticalQa}
  {\cal B}_{a}(t;\{\bfalp\}) =
  B_a(t;\{\alpha\}) - B'_a(t;\{\alpha'\}),
\ee
and
\be  \label{tilQIF}    \eqalign{
 B_a(t;\{\alpha\}) &\equiv\sum_{b} \int_{t_0}^{t}\!d\tau\,
 C_{ab}(t-\tau) Q_{b}[\alpha(\tau)], \cr
 B'_a(t;\{\alpha'\}) &\equiv\sum_{b} \int_{t_0}^{t}\!d\tau\,
 C^{\ast}_{ab}(t-\tau) Q_{b}[\alpha'(\tau)].}
\ee
Here, $C_{ab}(t)$ is the bath correlation functions, defined by \Eq{FFCorr}.
  The functional ${\cal A}_a$ [\Eq{calQAB}] depends
only on the local time and
its operator level form is just the commutator of
the dissipative mode $Q_a$.

  The functional ${\cal B}_a$ [\Eq{ticalQa}] does however contain memory,
which is to be resolved
via the COPI algorithm
of consecutive time derivatives on all memory--contained functionals.
To construct a close set of HEOM via the  COPI algebra,
it needs a
proper expansion of $C_{ab}(t)$ such as the exponential series, while maintaining
the fluctuation--dissipation theorem of \Eq{FDT}.
A super--Drude parametrization scheme and the
resulted HEOM  formalism have been presented in our previous work
\cite{Xu07031107}.
A hybrid scheme,  exploiting the analytical continuation
and quadrature integration methods to evaluate the fluctuation--dissipation
theorem, has also been proposed \cite{Zhe09164708}.

 To illustrate the COPI algorithm, consider the super-Drude model,
\be\label{Jmodelab}
  J_{ab}(\omega) = \frac{\eta_{ab}\omega
   +i\eta'_{ab}\omega^2}{[(\omega/\gamma_{ab})^2+1]^2} \, .
\ee
The parameters are all real and satisfy the symmetry relations
of $\eta_{ab}=\eta_{ba},\ \gamma_{ab}=\gamma_{ba}$, and
$\eta'_{ab}=-\eta'_{ba}$, as inferred from
the symmetric relations implied in $J_{ab}(\omega)$.
The resulting correlation functions via \Eq{FDT} are
\be\label{Cabtexpan}
  C_{ab}(t\geq 0)=[(\nu^{ab}_r+i\nu^{ab}_i)   
               + (\bar\nu^{ab}_r+i\bar\nu^{ab}_i)\, \gamma_{ab}t] e^{-\gamma_{ab}\,t}
               + \sum_{m=1}^{M} \check\nu_m^{ab} e^{-\check\gamma_mt}+\delta C_{ab}(t).
\ee
The second term arises from the Matsubara poles, with
$\check\gamma_m = 2\pi m/\beta$ the Matsubara frequencies,
and $\check\nu^{ab}_m=-i(2/\beta)J_{ab}(-i\check\gamma_m)=(\nu^{ab}_m)^\ast$
is real, as inferred from the symmetry relation of the spectral density function
in analytical continuation.
The first term arises from pole of the spectral density function of rank 2.
The involving coefficients are summarized as follows \cite{Yan05187,Xu07031107,Xu037}.
\be\label{nucoefapp}
 \eqalign{
  \nu^{ab}_i = \eta'_{ab}\gamma_{ab}^3/4,
 \quad    \bar\nu^{ab}_i = - {\frac{1}{4}}(\eta_{ab}
          +\eta'_{ab}\gamma_{ab})\gamma_{ab}^2\, ,
 \quad
       \bar\nu^{ab}_r = - \bar\nu^{ab}_i
       \cot(\beta\gamma_{ab}/2),
\cr
   \nu^{ab}_r
  = -\nu^{ab}_i \cot(\beta\gamma_{ab}/2)
    -\bar\nu^{ab}_i  (\beta\gamma_{ab}/2)
     \csc^2(\beta\gamma_{ab}/2)\, ,
}
\ee
and
\be \label{checknu}
  \check\nu^{ab}_m=-\frac{2(\eta_{ab}\check\gamma_m
        +\eta'_{ab}\check\gamma_m^2)}
            {\beta[(\check\gamma_m/\gamma_{ab})^2-1]^2}
  \equiv \check\gamma^2_m\check\eta^{ab}_m\, ,
  {\rm\ with\ \ }
   \check\eta^{ab}_m=-\frac{(\eta_{ab}+\eta'_{ab}\check\gamma_m)/(m\pi)}
     {[(\check\gamma_m/\gamma_{ab})^2-1]^{2}} .
\ee
The residue $\delta C_{ab}(t)$ can be
approximated via the Markovian ansatz,
i.e., $\delta C_{ab}(t)\simeq\Delta_{ab}\delta(t)$;  cf.\ \Eq{detC}.

 To proceed we denote for every distinct exponent terms in \Eq{Cabtexpan},
\be
\eqalign{
&
     B_{ab}(t;\{\alpha\}) \equiv
     \int_{t_0}^{t}\!d\tau\,
     e^{-\gamma_{ab}(t-\tau)} Q_b[\alpha(\tau)],
\cr
&
     \bar B_{ab}(t;\{\alpha\})  \equiv
     \int_{t_0}^{t}\!d\tau\, \gamma_{ab}(t-\tau)
     e^{-\gamma_{ab}(t-\tau)}   Q_b[\alpha(\tau)],
\cr
&
     \check B^{a}_{m}(t;\{\alpha\}) \equiv \sum_b \check\eta^{ab}_m
     \int_{t_0}^{t}\!d\tau\,
     e^{-\check\gamma_m(t-\tau)} Q_b[\alpha(\tau)];  \quad m=1,\cdots,M.}
\ee
They are related to the influence generating functionals as
[cf.\ \Eq{tilQIF}]
\be
\eqalign{
&         {\cal B}_{ab}   \equiv   -i(B_{ab}  -  B'_{ab}),
\qquad    {\cal B}'_{ab}   \equiv   B_{ab}   + B'_{ab},
\cr
&         \bar{\cal B}_{ab}  \equiv   -i({\bar B}_{ab}- {\bar B}'_{ab}),
\qquad    \bar{\cal B}'_{ab}  \equiv   {\bar B}_{ab}+ {\bar B}'_{ab},
\cr
&         \check{\cal B}^a_m   \equiv  -i(\check B^a_m  - \check B^{\prime a}_m);
\qquad\ \ m=1,\cdots,M.}
\ee
Now we define the auxiliary density operators (ADOs)
via \be \rho_{\sf n}(t)  \equiv   {\cal U}_{\sf n}(t,t_0)\rho(t_0),
\ee where
 \be \label{calGPIaux}
   {\cal U}_{\sf n}(\bfalp,t;\bfalp_0,t_0)
  \equiv \int_{\bfalp_0[t_0]}^{\bfalp[t]}   \!\!  {\cal D}{\bfalp} \,
     e^{iS[\alpha]} {\cal F}_{\sf n}[\bfalp] e^{-iS[\alpha']},
 \ee
with
\be\label{calFsfn}
 {\cal F}_{\sf n} =s_{\sf n}\Bigl\{\prod_{a,b}\bigl[
    ({\cal B}_{ab})^{n_{ab}}
    ({\cal B}'_{ab})^{n'_{ab}}
    (\bar{\cal B}_{ab})^{\bar n_{ab}}
    (\bar{\cal B}'_{ab})^{\bar{n}'_{ab}}\bigr]
    \prod_{a,m}\bigl(\check{\cal B}^{a}_{m}\bigr)^{\check n_{m}^{a}}
    \Bigr\} {\cal F}.
\ee
The scaling factor $s_{\sf n}$ is defined as
\be\label{ssfnapp}
 s_{\ind} =  \Big\{  \prod_{a,b}  
 \frac{
        |\nu_r^{ab}|^{n_{ab}}
        |\nu_i^{ab}|^{n'_{ab}}
        |\bar\nu_r^{ab}|^{\bar n_{ab}}
        |\bar\nu_i^{ab}|^{\bar n'_{ab}}}
   {    {n_{ab}}!\,
        {n'_{ab}}!\,
        {\bar n_{ab}}!\,
        {\bar n'_{ab}}!\,
   }  
   \Big\}^{\frac{1}{2}}
   \prod_{a,m}\frac{\check\gamma_m^{\check n_m^a}}
                             {\sqrt{\check n_m^a!}}.
\ee
The ADOs $\{\rho_{\sf n}\}$ defined in this way are dimensionless
and possess uniform error tolerance to support the filtering method,
as described in \Sec{theoryC}.
Since $\eta'_{aa}=0$ hence $\nu^{aa}_i=0$, we set $\nu^{aa}_i\equiv\nu^{aa}_r$
for the scaling factor $s_{\ind}$ of \Eq{ssfnapp} and hereafter.
The index set for this complex multi-dissipative modes case is now
\be \label{indnapp}
 {\ind}=
   \left\{n_{ab},n'_{ab},\bar n_{ab},\bar n'_{ab},
   \check n_1^a,\cdots,\check n_M^a \right\}.
\ee

  The HEOM can now
 be derived by taking the time derivative
 on $\rho_{\ind}$. Note that the COPI method is not just a time derivative technique,
 but provide a way to resolve the bath memory effect of the influence functional
 in the operative level.
 The final results are
\be \label{appdotrhon}
  \dot\rho_{\ind} = - [ i{\cal L}(t) + {\Gamma_{\ind}} + \delta\mathcal{R} ] \rho_{\ind}
       + \rhonswap + \rhondown + \rhonup ,
\ee
where
\bea
   {\Gamma}_{\ind}\equiv\sum_{a,b}\left( n_{ab}+{\bar n}_{ab}+n'_{ab}+{\bar n}'_{ab} \right) \gamma_{ab}
     +  \sum_{a,m} \check n^a_m \check \gamma_m   \,  ,
\label{appGamind} \\
\fl\qquad
  \delta\mathcal{R}\hat O = \sum_{a,b} \Delta_{ab} [Q_a,[Q_b,\hat O]].
\label{appdRinHEOM}
\eea
The details of the swap term $\rhonswap$,
the tier-down $\rhondown$ and tier-up $\rhonup$ terms are
\be
\rhonswap = \sum_{a,b} \gamma_{ab} \Bigl[
     \sqrt{(n_{ab}+1)  \bar n_{ab}  |\bar\nu_r^{ab}/\nu_r^{ab}|} \  {\rho}_{\vec{\ind}_{ab}}
   + \sqrt{(n'_{ab}+1) \bar n'_{ab} |\bar\nu_i^{ab}/\nu_i^{ab}|} \  {\rho}_{\vec{\ind}'_{ab}} \Bigr],
\ee
\bea
\rhondown =-i \sum_{a,b}    \sqrt{ n_{ab}|\nu_r^{ab}|}
              \left[ Q_b\,, {\rho}_{{\ind}_{ab}^{-}}\right]
            + \sum_{a,b}    \sqrt{n'_{ab}|\nu_i^{ab}|}
              \left\{Q_b\,, {\rho}_{{\ind}_{ab}^{\prime-}}\right\}
\nl \fl\qquad\ \qquad
           -i \sum_a        \sum_{m=1}^M  \sqrt{\check n_m^a}  \   \check\gamma_m
              \sum_b        \check\eta^{ab}_m
              \left[ Q_b\,, {\rho}_{{\check{\ind}}_{a,m}^{-}}\right],
\eea
\bea\label{appup}
\rhonup =
  -i \sum_{a,b}
  \left\{  \nu_r^{ab} \sqrt{(n_{ab}+1)  / |\nu_r^{ab}|}
           \left[Q_a\,,   {\rho}_{{\ind}^{+}_{ab}}  \right]
       +   \bar\nu_r^{ab}   \sqrt{(\bar n_{ab}+1)   /|\bar\nu_r^{ab}|}
           \left[Q_a\,,   {\rho}_{\bar{\ind}^+_{ab}}\right]
  \right.
\nl    \left.      \qquad           \qquad
      +   \bar\nu_i^{ab}   \sqrt{(\bar n'_{ab}+1)  /|\bar\nu_i^{ab}|}
           \left[Q_a\,,   {\rho}_{\bar{\ind}^{\prime+}_{ab}}\right]
       \right\}
\nl \fl  \qquad\ \qquad   -i \sum_{a\neq b}     \nu_i^{ab} \sqrt{(n'_{ab}+1) / |\nu_i^{ab}|}
           \left[Q_a\,,   {\rho}_{{\ind}^{\prime+}_{ab}}\right]
\nl \fl  \qquad\ \qquad   -i \sum_{a}\sum_{m=1}^{M} \sqrt{\check n_m^a+1}\ \check\gamma_m
           \left[Q_a\,,   {\rho}_{{\check{\ind}}^{+}_{a,m}}\right].
\eea
The index variations here are similar to those described in \Eq{sfnvar},
which is just the single--mode simplification.
Equation (\ref{rhonasso}) is thus obtained readily.

\section*{References}


\end{document}